\documentclass[sigconf]{acmart}

\usepackage{balance}

\AtBeginDocument{%
  }

\copyrightyear{2027}
\acmYear{2027}
\setcopyright{cc}
\acmDOI{XXXXXXX.XXXXXXX}
\acmISBN{978-1-4503-XXXX-X/2027/06}

\usepackage{amsmath}
\usepackage{amsthm} 
\usepackage[linesnumbered,ruled,vlined]{algorithm2e}
\newtheorem{theorem}{Theorem}

\usepackage[utf8]{inputenc}

\begin{document}

\title{Trust Matters: Unbiased Interleaving for Ranker Selection Under User Trust Bias}

\author{Dario Simionato}
\authornote{Corresponding author.}
\email{dario.simionato1@h-partners.com}
\affiliation{%
  \institution{Huawei Ireland Research Centre}
  \city{Dublin}
  \country{Ireland}
}

\author{Andrea Tonon}
\email{andrea.tonon1@huawei-partners.com}
\affiliation{%
  \institution{Huawei Ireland Research Centre}
  \city{Dublin}
  \country{Ireland}
}

\author{Mingxue Wang}
\email{wangmingxue1@huawei.com}
\affiliation{%
  \institution{Huawei Ireland Research Centre}
  \city{Dublin}
  \country{Ireland}
}

\author{Xiaoyue Li}
\email{lixiaoyue4@huawei.com}
\affiliation{%
  \institution{Huawei Nanjing R\&D Center}
  \city{Nanjing}
  \country{China}
}

\author{Tong Gui}
\email{guitong@huawei.com}
\affiliation{%
  \institution{Huawei Nanjing R\&D Center}
  \city{Nanjing}
  \country{China}
}

\author{Xijian Ju}
\email{juxijian@huawei.com}
\affiliation{%
  \institution{Huawei Nanjing R\&D Center}
  \city{Nanjing}
  \country{China}
}

\renewcommand{\shortauthors}{Dario Simionato et al.}


\begin{abstract}

Interleaving experiments are widely used in place of A/B tests to accelerate comparisons between ranking systems. 
Yet, despite their theoretical guarantees, existing "unbiased" interleaving methods can produce misleading conclusions in practice because their analyses rely on simplified user behavior assumptions that do not hold in real-world settings. 
As a result, an important limitation of existing methods has remained largely unrecognized: their guarantees can fail under realistic forms of user behavior such as trust bias, where users develop trust in ranking systems and preferentially engage with higher-ranked results. 
Despite a large body of work on interleaving, the field lacks a principled framework that accounts for this behavior.

To resolve this fundamental limitation, we introduce \textit{Trust-Aware Balanced Interleaving}, the first interleaving framework that provides provably unbiased ranker comparisons and fidelity under arbitrary levels of trust bias through a novel credit-attribution mechanism.
Building on this framework, we show that existing propensity-scored interleaving methods implicitly rely on positivity assumptions that, if violated under trust bias, lead to uncontrolled false positive outcomes revealing a previously unrecognized failure of such methods.
We further show that standard propensity-scored interleaving emerges as a special case of our framework under the restrictive assumption of no trust bias, thereby connecting prior work to a more general theoretical formulation.

We then study whether our unbiased interleaving can be extended to multileaving settings and show, through a counterexample, that naive generalizations fail to preserve fidelity.
This lead us to develop an efficient pairwise comparisons framework that enables unbiased candidate selection without requiring quadratic comparisons, allowing our framework to identify a set of top-$k$ rankers for subsequent A/B/n testing.
Experimental evaluations demonstrate that our approaches provide unbiased ranker comparisons under trust bias and enable reliable ranker selection while retaining the sample efficiency of interleaving.
\end{abstract}



\maketitle

\section{Introduction}\label{sec:intro}

\begin{figure}[t]
\centering
\begin{tabular}{llllllll}
\textcolor{blue}{Ranker $\mathcal{R}_1$} &
$i_1$ & $i_2$ & $i_3$ & $i_4$ & $i_5$ & $i_6$ & $i_0$ \\
\textcolor{red}{Ranker $\mathcal{R}_2$} &
$i_5$ & $i_1$ & $i_2$ & $i_3$ & $i_4$ & $i_0$ & $i_6$ \\
\hline
Interleave $o=(\mathcal{R}_1,\mathcal{R}_2)$ &
\textcolor{blue}{$i_1$} &
\textcolor{red}{$i_5$} &
\textcolor{blue}{$i_2$} &
\textcolor{blue}{$i_3$} &
\textcolor{blue}{$i_4$} &
\textcolor{blue}{$i_6$} &
\textcolor{red}{$i_0$}
\\
Interleave $o=(\mathcal{R}_2,\mathcal{R}_1)$ &
\textcolor{red}{$i_5$} &
\textcolor{blue}{$i_1$} &
\textcolor{blue}{$i_2$} &
\textcolor{blue}{$i_3$} &
\textcolor{blue}{$i_4$} &
\textcolor{red}{$i_0$} &
\textcolor{blue}{$i_6$}
\end{tabular}
\caption{
Two ranked lists (top) and the two possible resulting interleaved lists (bottom).
The interleaving assigns unequal numbers of slots to each ranker: Ranker $\mathcal{R}_1$ contributes 5 items (blue), while Ranker $\mathcal{R}_2$ contributes 2 items (red).
}\label{fig:ex_1}
\end{figure}

Online experimentation plays a crucial role in the development of ranking systems. 
The standard approach is A/B testing, where users are exposed to one ranking system at a time and differences are measured through online metrics such as clicks or conversions. 
While reliable, A/B testing often requires substantial traffic before reaching statistically significant conclusions, especially if more than two versions are compared. 
To accelerate experimentation, interleaving methods, which compare multiple ranking systems within the same user interaction interface and achieve significantly lower variance and faster decisions, have been proposed.

The increased efficiency of interleaving comes at the cost of a more complex evaluation. 
In an A/B test, users interact with complete rankings produced by a single system. 
In contrast, interleaving constructs mixed rankings by combining results from competing systems and then infers preferences from the observed interactions.
As a result, interleaving outcomes are harder to interpret and may diverge from A/B test results, complicating failure analysis.

Over the years, several interleaving methods have been proposed, targeting properties such as fidelity (i.e., detecting the best ranker under infinite user feedback) and unbiasedness (i.e., not detecting any best ranker under no user preference), whose theoretical guarantees are based on simplistic click models of user behavior~\cite{chapelle2012large}. 
Among these, the work of Amazon~\cite{bi2022debiasedinterleaving} is the most principled and practically applicable approach, providing provable unbiasedness guarantees and fidelity under an inverse propensity scoring framework.
The key idea is to reweight the credit assigned to each ranker according to its probability of receiving a click, thereby compensating for the unequal representation of competing rankers in the interleaved list (Figure~\ref{fig:ex_1}).
These guarantees, however, rely on the simplifying assumption that users click viewed items uniformly at random, implying that all displayed positions are equally likely to receive a click and therefore contribute equally to the evaluation.

Such random clicking behavior, however, is not representative of real-world scenarios where users tend to preferentially click higher-ranked results rather than lower-ranked ones, as they develop trust in the ranking system’s ability to surface high-quality items at the top of the list. 
We refer to this phenomenon as \emph{trust bias}~\cite{ovaisi2024fairness, agarwal2019addressing}, which is pervasive as real-world systems improve over time surfacing more results that match user needs. 
While recent work has shown that ignoring this behavior can invalidate evaluation estimators in related ranking paradigms~\cite{vardasbi2020when}, existing online interleaving methods provide guarantees only under idealized random-click assumptions.
This issue arises in interleaved rankings as well: under trust bias clicks change across positions implying that they should contribute differently to the comparison.

Beyond a single pairwise comparison, practical systems must often evaluate and select among many candidate ranking functions. 
Prior work has extended interleaving to multiple rankers (known as multileaving), but these approaches typically neither provide provable unbiasedness guarantees nor account for trust bias.
Moreover, they require generating more than two ranked lists per query and do not support caching, potentially increasing user latency and limiting their applicability for online deployment.
An alternative standard strategy is to aggregate interleaving results based on exhaustive pairwise comparisons~\cite{bi2023multipleinterleaving}, where each pair of systems is evaluated independently and the results are combined to determine an overall winner. 
While conceptually simple, this strategy scales quadratically in the number of systems and may lead to selection outcomes that do not necessarily align with the system that would perform best under a full A/B test. 
This bottleneck becomes critical in large-scale industrial settings where many rankers must be compared efficiently and reliably.

In this paper we resolve the fundamental limitation of interleaving methods not providing theoretical guarantees under the realistic click models of users under trust bias while retaining fidelity. 
We introduce \textit{Trust-Aware Balanced Interleaving}, the first interleaving framework that provides provably unbiased comparisons under arbitrary levels of trust bias by explicitly incorporating trust bias through a novel credit-attribution mechanism.
Building on this formulation, we show that existing propensity-scored interleaving methods can be recovered as a special case of our framework under the assumption of no trust bias.
We further demonstrate that these methods implicitly rely on positivity assumptions that, if not guaranteed under realistic trust-biased interactions, lead to biased estimates and uncontrolled false positive outcomes. 
This reveals a previously unrecognized failure mode of state-of-the-art “unbiased” interleaving methods when deployed in realistic settings, and provides a unified lens through which prior work can be understood as a restricted instance of our more general formulation.
To the best of our knowledge, we are the first to provide a theoretical analysis of interleaving under trust-biased user behavior, and to design an interleaving framework with provable guarantees in this setting.

We then study whether our algorithm can be extended to multileaving settings and show, through a counterexample, that a direct extension fails to preserve fidelity, posing challenges to efficient and unbiased selection among multiple rankers under trust bias.
We therefore introduce a new method for aggregating multiple (pairwise) interleaving comparisons that identifies a subset of top-$k$ candidate rankers requiring fewer comparisons than exhaustive pairwise evaluation.
The resulting method is simple, interpretable, and easy to debug, making it suitable for industrial deployment.

Overall, our work bridges the gap between theoretical guarantees and practical deployment by enabling fast, simple, and provably unbiased interleaving evaluation under realistic user behavior.

\section{Related work}\label{sec:rel_work}
User feedback has long been a central signal for evaluating and improving ranking systems in online settings. 
However, clicks are known to be biased, most notably due to position bias, where items ranked higher receive more clicks simply because they are more likely to be examined by users, rather than because they are more relevant~\cite{craswell2008experimental, richardson2007predicting,ovaisi2024fairness}. 
Beyond this, trust bias further skews user behavior, where users implicitly trust the system's ordering and preferentially click higher-ranked results because perceived as more relevant~\cite{joachims2005accurately, agarwal2019addressing, ovaisi2024fairness}. 
\citet{vardasbi2020when} showed that trust bias invalidates inverse propensity scoring for counterfactual learning-to-rank and proposed an affine correction to estimate relevance from biased clicks. 
Our work addresses a different problem: unbiased online comparison of ranking systems through interleaving, where clicks are attributed to competing rankers rather than used to estimate relevance. 
Consequently, their estimator is not directly applicable, and new mechanisms are required to preserve unbiasedness in the interleaving setting.
Because the implications of trust bias for interleaving-based online evaluation remain largely unexplored, this work introduces the first framework to study interleaving under trust-biased user behavior, deriving algorithms with rigorous theoretical guarantees in such scenario.

Interleaving methods aim to provide a more sample-efficient alternative to A/B testing for online evaluation of ranking systems by combining two candidate lists ($L_1, L_2$) into a single merged ranking ($I$) and inferring preferences from user clicks.
The choice of merging policy and credit assignment determines an algorithm's practical bias, i.e. whether it can produce incorrect winner selections, and fidelity, i.e. whether it has no power (the ability to detect the best ranker, if any) for specific ranker list-user interaction pairs~\cite{chapelle2012large}. 
Over the years, interleaving methods have been developed primarily to achieve unbiasedness and fidelity, while still largely overlooking the presence of trust bias.
The first interleaving approach is balanced interleaving (BI)~\cite{joachims2002optimizing,joachims2003evaluating}, which constructs the interleaved list by alternating picks between the two rankers' lists. 
If a ranker proposes an item already selected, it forfeits its turn. 
For credit assignment, a ranker receives credit based on how many clicked items originally appeared within its top-$k$ elements.
While highly efficient, BI suffers from alignment bias: when lists are nearly identical, minimal item shifts systematically distort credit allocation.
This results in BI being biased under random user clicks, favoring one ranker over the other even when both are equally effective and leading to false positives in output.
Team Draft Interleaving (TDI)~\cite{radlinski2008tdi}, which mimics a sports draft, was proposed to solve such issue. 
For each position in $I$, a coin toss assigns priority to a ranker, which then "drafts" its highest-ranked unselected item. 
Clicks are mapped directly back to the drafting ranker, ensuring an intuitive, unbiased assignment.
The main drawback of TDI lies in its violation of fidelity: under certain ranker assignments and click profiles, it returns a tie between two rankers even when one is in fact the clear winner hence resulting in a false negative.
Such limitation has been overcome by Probabilistic Interleaving~\cite{hofmann2011probabilistic} and Optimized Interleaving~\cite{radlinski2013optimized}, which either take into account all possible combinations of items sampled from the two ranked lists or solve an optimization problem to produce interleaved rankings that maximize the ability to distinguish between competing rankers.
Such approaches, despite their favorable statistical properties, are not widely adopted in practice due to implementation and computational constraints~\cite{zhang2025airbnbTD}.

Recent papers advanced the previous versions making them industry-ready, of which two are worth noticing.
\citet{zhang2025airbnbTD} extended TDI assignment and credit attribution with the concept of competitive pairs: if at any step an item is suggested by both rankers, it is added to the list without any assigned team and ignored from the analysis. 
Such tweak improved the power of TDI but did not solve its lack of fidelity.
\citet{bi2022debiasedinterleaving} proposed a debiased variant of BI and proved its unbiasedness under a random-click model. 
The method preserves BI's balanced merging strategy, which has the practical advantage of producing only two possible interleaved rankings for any pair of rankers, allowing the interleaving step to be precomputed and cached.
In addition, it introduces a revised credit attribution scheme that corrects for the bias caused by highly overlapping rankings.
When both rankers propose the same item, the item is inserted into the interleaved list but excluded from the credit computation.
Credit assignment is then adjusted using inverse propensity weighting to account for unequal exposure in the interleaved list, thus correcting for situations in which one ranker receives more click opportunities than the other (as in Figure~\ref{fig:ex_1}).
However, this correction is valid only under a simplified assumption of random user clicks, with no dependence on item position. 
In this work, we introduce a new interleaving method that remains unbiased and satisfies fidelity in more realistic settings where user clicks are influenced by trust bias, while retaining caching capabilities.
We further show that~\citet{bi2022debiasedinterleaving} is recovered as a special case of our framework when trust bias is absent, allowing us to characterize the failure modes of existing inverse-propensity interleaving methods in such settings.

Several works have extended interleaving to compare more than two rankers simultaneously~\cite{schuth2014multileaved,schuth2015probabilistic}, giving rise to the so-called multileaving algorithms, which aim to infer a relative ordering among multiple ranking systems.
While this additional flexibility improves experimental efficiency, it also complicates both failure analysis and bias correction~\cite{bi2022debiasedinterleaving}. 
In particular, while some prior work on multileaving provides provable unbiasedness guarantees~\cite{brost2016SOSM, oosterhuis2017ppm}, their proofs do not account for trust bias under realistic user-behavior models.
Crucially, even if a trust-bias correction were mathematically derived for these frameworks, their procedures produce superlinear sets of possible interleaved lists, thereby ruling out the practical offline caching needed for web-scale production deployment, as users would otherwise observe different rankings upon refreshing the page or submitting the same query again.
The other major obstacle to deploying multileaving in production is in fact the increased computational overhead, as more than two rankings must be generated for all candidate systems being evaluated, possibly causing costly delays on the user-side.
To mitigate this issue while still identifying a subset of promising candidates for subsequent A/B/n testing,~\citet{bi2023multipleinterleaving} propose performing pairwise interleaving experiments between all ranker pairs and then aggregating the outcomes into a global ranking.
However, this approach requires a number of comparisons that grows quadratically with the number of rankers and relies on the assumption of pairwise interleaving preferences to be transitive, which may not hold in practice.
Another line of work considers more complex selection strategies based on multi-armed bandit formulations and related sequential decision-making methods~\cite{li2020mergedts,zoghi2015mergerucb}. 
While these approaches can improve exploration efficiency, they typically rely on additional structural assumptions and introduce significant algorithmic complexity, making them less suitable for the interleaving setting considered here.
In this work, we present an iterative candidate selection method that performs repeated pairwise interleaving in a Swiss-system tournament style while progressively pruning low-performing candidates to reduce computational cost. 
Our algorithm achieves sub-quadratic complexity and returns the set of top-$k$ rankers without requiring additional assumptions such as the transitivity of pairwise interleaving preferences.

\section{Preliminaries}\label{sec:preliminaries}
We consider a pairwise interleaving setting\footnote{The setup extends naturally to multileaving by considering additional rankers and sampling from all permutations of the participating rankers.} with rankers $\mathcal{R}_1$ and $\mathcal{R}_2$ that produce ordered lists $L_1=\langle i_{1,L_1},i_{2,L_1},\ldots,i_{\ell,L_1}\rangle$ and $L_2=\langle i_{1,L_2},i_{2,L_2},\ldots,i_{\ell,L_2}\rangle$ of length $\ell$. 
Given $(L_1,L_2)$, an interleaving algorithm $\mathcal{I}$ samples a random ordering $o\in\{(\mathcal{R}_1,\mathcal{R}_2),(\mathcal{R}_2,\mathcal{R}_1)\}$ and produces an interleaved ranking $I=\langle i_1,i_2,\ldots,i_\ell\rangle$ containing all items passed in input. 
In addition, $\mathcal{I}$ defines a credit-assignment function $A(i)\in\{\mathcal{R}_1,\mathcal{R}_2,\emptyset\}$ specifying which ranker receives credit (if any) if item $i$ is clicked. 
We use $\mathbf{1}(A(i_k) = \mathcal{R})$ as an indicator function being 1 if $A(i_k) = \mathcal{R}$ and 0 otherwise.

The just-defined parameters refer to one single instantiation of the interleaving algorithm, but its performances are to be calculated modeling user behaviors and thus requiring additional notations. 
Following~\citet{bi2022debiasedinterleaving}, we define $V(i)$ and $E(i)$ as the binary variables indicating if the user views and engages (e.g. clicks) with item $i$, respectively.
We introduce a rank-dependent trust bias $b(k)\in[0,1]$~\cite{agarwal2019addressing}, which models the effect of rank on users' perceived relevance, making higher-ranked items appear more relevant than lower-ranked ones. 
Consequently, even after viewing all items, users are more likely to engage with items appearing higher in the ranking than with equally relevant items appearing lower down.
This translates into modeling the click probability given the user observes an item $i_k$ at position $k$ as $P(E(i_k)=1|V(i_k)=1) = rel_p(i_k)$ where the perceived relevance $rel_p(i_k)=b(k) \cdot rel(i_k)$ combines the item's intrinsic relevance $rel(i_k)$ with a trust bias $b(k)$~\cite{ovaisi2024fairness}.

\section{A Generalized Framework for Credit Attribution Under Trust Bias}
\label{sec:trust_aware_interleaving}

To evaluate rankers in online experiments, an evaluation framework is needed in which credit attribution reflects users’ intrinsic preferences rather than biases induced by trust in higher-ranked results. 
In this section, we introduce \textit{Trust-Aware Balanced Interleaving}, a generalized framework designed to correct for trust bias in credit attribution. 
We first define our credit-attribution function from first principles, then show that prior propensity-scored methods arise as a restricted case of our formulation, and characterize the failure modes that emerge when this assumption is violated.

\subsection{Trust-Aware Balanced Interleaving}
To achieve provable unbiasedness in environments governed by position-dependent user trust, we introduce our \textit{Trust-Aware Balanced Interleaving}.
Our method follows the balanced interleaving procedure for constructing interleaved rankings~\cite{joachims2002optimizing,joachims2003evaluating} and assigns credit to a ranker only when an interacted item $i$ is uniquely proposed by that ranker, excluding items suggested by both competing rankers~\cite{bi2022debiasedinterleaving,zhang2025airbnbTD}.
Our method uses a corrected credit metric $\tilde{C}_{\mathcal{R}}^{corr}(I)$ that inversely weights user engagement on viewed items by a rank-dependent trust-aware correction factor $\hat{d}_{\mathcal{R}}^{corr}$. Formally, for a ranker $\mathcal{R}$, the corrected attribution is defined as:
\begin{equation}\label{eq:our_credit_computation}
\tilde{C}_{\mathcal{R}}^{corr}(I) = \sum_{i_k \in I} \frac{\mathbf{1}(A(i_k) = \mathcal{R}) V(i_k) E(i_k)}{\hat{d}_{\mathcal{R}}^{corr}}
\end{equation}
where the denominator $\hat{d}_{\mathcal{R}}^{corr}$ weights $\mathcal{R}$ click opportunities using the trust bias $b(k)$ measured on the set of viewed slots:
\begin{equation}
\hat{d}_{\mathcal{R}}^{corr} = \frac{\sum_{i_k \in I} \mathbf{1}(A(i_k) = \mathcal{R}) V(i_k) b(k)}{\sum_{i_k \in I} \mathbf{1}(A(i_k) \neq \emptyset) V(i_k) b(k)}
\end{equation}
By computing the propensity weights as a function of $b(k)$, this formulation corrects for the tendency of users to trust higher-ranked items more than lower-ranked ones, ensuring that credit attribution reflects intrinsic relevance rather than positional trust effects.

To formally characterize this guarantee, we define a \textit{uniform-relevance user} as an uninformed user considering documents relevance to be constant across all items (i.e., $rel(i) = \alpha$ for all $i$). 
Importantly, user clicks are still governed by trust bias in the examination process.
Under this condition, an unbiased interleaving algorithm should result in a tie in expectation, as the user is not intentionally expressing any preference between the two rankers.
We theoretically show that Trust-Aware Balanced Interleaving satisfies this property as follows:
\begin{theorem}
For a uniform-relevance user under trust bias $b$, the Trust-Aware Balanced Interleaving returns a tie in expectation:
\[\mathbb{E}_{o,V,E}[\tilde{C}^{corr}_{\mathcal{R}_1}(I) - \tilde{C}^{corr}_{\mathcal{R}_2}(I)] = 0\] 
\end{theorem}
\begin{proof}
For any given user engagement with the interleaved list, let $K_V = \{k \mid V(i_k)=1\}$ be the set of viewed slots. 
In such user engagement case, we got $\hat d^{corr}_\mathcal{R}=\frac{\sum_{k\in K_V}\mathbf{1}(A(i_k)=\mathcal{R})b(k)}{\sum_{k\in K_V}\mathbf{1}(A(i_k)\not=\emptyset)b(k)}$.

By the law of total expectation, conditioning on the interleaving assignment $o$ and the viewed slots $K_V$, the conditional expectation of credit for ranker $\mathcal{R} \in \{\mathcal{R}_1, \mathcal{R}_2\}$ becomes:
\begin{align}
\mathbb{E}_{E \mid o, K_V}&\left[ \tilde{C}^{corr}_\mathcal{R}(I) \right]= \nonumber\\
&= \sum_{k \in K_V} \frac{\mathbf{1}(A(i_k)=\mathcal{R}) \cdot \mathbb{E}_{E \mid o, K_V}[E(i_k)]}{\hat d^{corr}_\mathcal{R}}= \nonumber\\
&= \frac{\sum_{k \in K_V}\mathbf{1}(A(i_k)=\mathcal{R}) \cdot \mathbb{E}_{E \mid o, K_V}[E(i_k)]}{\frac{\sum_{k\in K_V}\mathbf{1}(A(i_k)=\mathcal{R})b(k)}{\sum_{k\in K_V}\mathbf{1}(A(i_k)\not=\emptyset)b(k)}}= \nonumber 
\end{align}
\begin{align}
&= \frac{\sum_{k \in K_V}\mathbf{1}(A(i_k)=\mathcal{R}) \cdot \alpha \cdot b(k)}{\frac{\sum_{k\in K_V}\mathbf{1}(A(i_k)=\mathcal{R})b(k)}{\sum_{k\in K_V}\mathbf{1}(A(i_k)\not=\emptyset)b(k)}}= \nonumber\\
&= \alpha \cdot \frac{\sum_{k \in K_V} \mathbf{1}(A(i_k)=\mathcal{R})b(k)}{\frac{\sum_{k \in K_V} \mathbf{1}(A(i_k)=\mathcal{R})b(k)}{\sum_{k \in K_V} \mathbf{1}(A(i_k)\neq\emptyset)b(k)}}= \nonumber \\
&= \alpha \cdot \sum_{k \in K_V} \mathbf{1}(A(i_k)\neq\emptyset)b(k) \nonumber 
\end{align}
The ranker-specific terms cleanly cancel out and the remaining sum is completely invariant to the ranker identity $\mathcal{R}$ leading to:
\begin{equation*}
\mathbb{E}_{E \mid o, K_V}\left[\tilde{C}^{corr}_{\mathcal{R}_1}(I)\right] - \mathbb{E}_{E \mid o, K_V}\left[\tilde{C}^{corr}_{\mathcal{R}_2}(I)\right] = 0
\end{equation*}
Applying the law of iterated expectations over all interleaving choices $o$ and visibility vectors $V$ yields the final result:
\begin{align}
\mathbb{E}_{o,V,E}&[\tilde{C}^{corr}_{\mathcal{R}_1}(I) - \tilde{C}^{corr}_{\mathcal{R}_2}(I)] = \nonumber\\ 
&=\mathbb{E}_{o,V}\left[ \mathbb{E}_{E \mid o, K_V}\left[\tilde{C}^{corr}_{\mathcal{R}_1}(I) - \tilde{C}^{corr}_{\mathcal{R}_2}(I)\right] \right] = \nonumber\\ 
& = \mathbb{E}_{o,V}[0] = 0.\nonumber
\end{align}
\end{proof}
We note that our credit assignment $\tilde{C}^{corr}_{\mathcal{R}_1}(I)$ is undefined if and only if $\hat d^{corr} = 0$. 
This boundary case occurs when no items within the user's viewed positions are attributed to any ranker ($\forall k \in K_V, A(i_k) = \emptyset$), which happens if and only if the participating rankers produce identical ordering configurations over those specific slots. 
In these instances, we define both assigned credits to be $0$, naturally resulting in a tie.
A Z-test is then applied to the total user credits to assess statistical differences between rankers.

Our attribution policy assigns credit for shared items exclusively to the ranker that places the item higher, and assigns no credit when positions are identical.
This is consistent with the fidelity definition~\cite{hofmann2011probabilistic,radlinski2013optimized}, which require that interactions on commonly ranked items are attributed to the higher-ranked system, thereby ensuring rank-sensitive credit assignment.

While the theoretical unbiasedness of Trust-Aware Balanced Interleaving relies on the true underlying trust bias $b(k)$, this function might be unknown in practice. 
To bridge this gap and deploy the framework in a live production environment, we estimate the trust bias asynchronously from user behavioral logs collected natively during the interleaving experiment alongside predicted document relevances. 
These relevance estimates are typically available as pre-click scores produced by the underlying ranking model, which may itself be debiased for trust bias using existing correction methods~\cite{agarwal2019addressing,vardasbi2020when} or obtained from external relevance proxies~\cite{chapelle2011yahoo, liu2009learning}.
Formally, let $N_{click}(k)$ denote the total number of clicks received by items displayed at position $k$ across all users and all experimental sessions. 
Furthermore, let $N_{exp}(k)= \sum_{i\in \mathcal{D}_k} rel(i)$ represent the expected number of clicks in the absence of trust
bias, where $\mathcal{D}_k$ is the set of all items displayed at position $k$ throughout the duration of the interleaved experiment, and $rel(i)$ is the inferred intrinsic relevance of item $i$. 
We then compute the empirical trust bias estimate at position $k$ as $\hat b(k)= N_{click}(k) / N_{exp}(k)$. 
Intuitively, $\hat b(k)$ measures the multiplicative deviation between the observed click frequency and the baseline frequency that would be expected purely from item relevance (hence without trust bias). 
By comparing the observed click frequency with the frequency expected from item relevance alone, $\hat b(k)$ isolates the effect difference attributable to position providing an estimate of the trust bias at index $k$.

\subsection{Reduction to Standard Inverse-Propensity Estimators}\label{sec:reduction}
Our formulation generalizes inverse-propensity interleaving methods to trust bias scenarios and reduces to classical balanced interleaving under a uniform trust assumption. In particular, the debiased balanced interleaving framework of Amazon~\cite{bi2022debiasedinterleaving} can be recovered as a special case. 

Assume that user trust does not depend on position, i.e., ($b(k) = \beta$) for all $k$. 
Under this condition, the trust-aware propensity term simplifies as follows:

\begin{equation*}
\hat{d}_{\mathcal{R}}^{\text{cor}} \Big|_{b(k)=\beta} = \frac{\sum_{i_k} \mathbf{1}(A(i_k) = \mathcal{R}) V(i_k) \beta}
{\sum_{i_k} \mathbf{1}(A(i_k) \neq \emptyset) V(i_k) \beta} 
= \frac{\sum_{i_k} \mathbf{1}(A(i_k) = \mathcal{R}) V(i_k)}
{\sum_{i_k} \mathbf{1}(A(i_k) \neq \emptyset) V(i_k)} .
\end{equation*}
This expression matches the standard empirical normalization used in prior balanced interleaving methods~\cite{bi2022debiasedinterleaving}. Thus, classical formulations implicitly correspond to the special case where trust is uniform across positions. Our framework therefore strictly generalizes these methods while remaining consistent with them under this limiting condition.

\subsection{Failure of Standard Inverse-Propensity Estimators under Trust Bias}
Standard inverse-propensity interleaving estimators can become biased due to violations of the positivity condition in presence of trust bias.
To illustrate this issue, consider the presentation layout in Figure~\ref{fig:ex_1}. 
In this configuration, items in slots 3 to 5 are deterministically assigned to Ranker $\mathcal{R}_1$, regardless of the interleaving randomization $o$. 
This violates the standard positivity requirement, which requires that every position have non-zero probability of being attributed to either ranker, i.e., $P(A(i_k)=\mathcal{R}) > 0$.
To overcome such imbalances, standard estimators use a global normalization over the entire list. 
However, this introduces a mismatch under position-dependent trust bias $b(k)$. 
In particular, Ranker $\mathcal{R}_1$ receives more attribution opportunities due to control over slots 3–5, and the standard estimator compensates by reducing its overall credit to correct for this imbalance.
This correction implicitly treats all positions as equally important. 
However, slots 3–5 lie in lower-ranked positions where user engagement might be influenced by non-constant trust bias $b(k)$. 
As a result, the estimator might wrongly quantify the correction for the additional exposure of $\mathcal{R}_1$, without accounting for the different effective interaction probability in these positions.

Consequently, considering a uniform-relevance user under trust bias, the baseline estimator of~\citet{bi2022debiasedinterleaving} assigns higher expected credit to $\mathcal{R}_2$. 
This discrepancy arises from treating exposure opportunities as equally valuable across positions, without incorporating trust-weighted engagement. 
As shown in Appendix~\ref{sec:app_ex_1}, this may lead to systematic distortion in the presence of trust bias, increasing the likelihood of incorrect preference attribution between rankers. 
Our trust-aware correction addresses this issue by weighting exposure propensities according to position-specific trust bias $b(k)$, aligning credit attribution with effective interaction probability.

\section{Multiple Rankers Selection under Trust Bias}
\subsection{Limitations of Extending to Multileaving}\label{sec:our_multileaving}

\begin{figure}[t]
\centering
\begin{tabular}{llll}
Ranker $\mathcal{R}_1$ &
$\mathbf{i_1}$ & $\mathbf{i_2}$ & $i_3$ \\
Ranker $\mathcal{R}_2$ &
$\mathbf{i_2}$ & $i_3$ & $\mathbf{i_1}$ \\
Ranker $\mathcal{R}_3$ &
$i_3$ & $\mathbf{i_1}$ & $\mathbf{i_2}$ \\
\end{tabular}
\caption{
Example with three rankers. Clicked items are shown in boldface.
}\label{fig:ex_2}
\end{figure}

Given its proven unbiasedness under trust bias, we investigate the extension of Trust-Aware Balanced Interleaving to a multileaving setting.
We sample a permutation $o$ over all participating rankers and redefine the credit assignment function $A$ such that $A(i)=\emptyset$ only if all rankers propose item $i$ otherwise credit is assigned to every ranker that proposes $i$.
Such design, however, scales poorly with the number of rankers in input as in the case of Figure~\ref{fig:ex_2}.
Consider a case where the user clicks on two items $i_1$ and $i_2$: the optimal ranker is $\mathcal{R}_1$, since it ranks both clicked items in the top positions. 
However, for any permutation $o$, a click on $i_1$ assigns credit exclusively to $\mathcal{R}_1$, while a click on $i_2$ assigns credit exclusively to $\mathcal{R}_2$. 
As a result, the interleaver incorrectly returns a tie between $\mathcal{R}_1$ and $\mathcal{R}_2$, failing to identify the true winner (i.e., violating fidelity). 
A similar limitation also arises when the number of rankers is larger than the number of ranked items, in which multiple rankers propose the same item as the top-ranked and each item is the first recommendation for at least one ranker. 
In such cases, the algorithm cannot reliably discriminate between rankers within each group sharing the same top suggestion, as credit is only attributed to that shared item.

\subsection{Iterative Candidate Selection via Pairwise Interleaving}\label{sec:iterative_selection}
Motivated by such limitations, we propose an iterative candidate selection strategy that identifies a subset of top-$k$ candidate rankers for subsequent A/B/n testing through multiple pairwise interleaving tests following a Swiss-tournament style.

We first notice how interleaving and A/B test results may differ~\cite{chapelle2012large} even for metrics that are positively correlated, meaning that an higher relevance is expected to yield both higher credits and higher A/B test metrics.
To illustrate this, consider the example of Figure~\ref{fig:ex_1} under the assumption that for each item $rel(i_x)=\frac{1}{1+x}$, and that the user examines all items in the ranked list.
As shown in Appendix~\ref{sec:app_ex_2}, our Trust-Aware Balanced Interleaving can assign higher expected credit to $\mathcal{R}_2$ than to $\mathcal{R}_1$, whereas an A/B test based on average clicks per session would rank $\mathcal{R}_1$ above $\mathcal{R}_2$.
These discrepancies arise from the fact that interleaving alters the positions at which items are displayed relative to the original rankings, leading to different exposure patterns under rank-dependent trust bias, which can affect interleaving and A/B evaluation differently.
We observe that such difference in outcomes depends both on the interleaved lists and the trust bias function $b$.

This observation motivates the design of an algorithm that evaluates each ranker multiple times, with the expectation that such discrepancies average out due to probabilistic effects (pseudocode in Appendix~\ref{sec:app_sel_pseudocode}). 
The algorithm takes as input a set of rankers $\mathbf{R}$, the desired number of output rankers $k$, the chosen interleaving method, a maximum number of rounds $T$ and a significance threshold $\alpha$ (global across the $T$ rounds). 
It returns a list $\mathbf{S}$ containing the top-$k$ rankers.
The method performs multiple comparisons following a Swiss-tournament style ordering, where at each iteration the current set of candidate rankers is randomly partitioned into pairs of comparable performances and evaluated via interleaving.
The outcome of each comparison is used to update per-ranker statistics capturing wins, ties, and losses. 
After each round, these statistics are aggregated into a score that reflects the current evidence of each ranker’s relative performance. 
Rankers that are statistically unlikely to belong to the top-$k$ set, or that are dominated in terms of accumulated wins, are progressively eliminated from the candidate pool.
This process is repeated for at most $T\ll N$ rounds but may terminate earlier if only $k$ candidates remain in the pool.
The final selection is obtained by ranking the surviving candidates according to their aggregated scores, with tie-breaking based on the strength of their opponents, and selecting the top-$k$.
At each round, our method performs at most $N/2$ interleaving comparisons, resulting in a worst case computational complexity of $\mathcal{O}(NT)$ which is strictly sub-quadratic in $N$.

Our candidate pruning either removes ranker that cannot make it in the top-$k$ in the remaining rounds or those that performs statistically worse than random without removing the true top-$k$ rankers as follows (proof in Appendix~\ref{sec:app_proof_selection}):
\begin{theorem}\label{thm:selection}
    The pruning mechanism of the Iterative Candidate Selection algorithm removes a ranker in the true set of top-$k$ rankers with probability at most $\alpha$.
\end{theorem}

Our algorithm assumes a positive correlation between A/B test metrics and interleaving-based credit attribution, which is in line with knowledge from the literature~\cite{bi2022debiasedinterleaving, bi2023multipleinterleaving, zhang2025airbnbTD}. 
The method can be instantiated with any unbiased interleaving scheme, making it broadly applicable, including in more complex future scenarios.

\section{Experimental Evaluation}\label{sec:experiments}

We evaluate our proposed methods in two complementary experiments\footnote{The implementation code is available in an anonymized repository: \url{https://anonymous.4open.science/r/TrustAwareBalancedInterleaving/}.}. 
Section~\ref{sec:tabi-eval} evaluates Trust-Aware Balanced Interleaving by measuring its ability to control bias and preserve statistical power under trust-bias, compared to existing interleaving methods.
Section~\ref{sec:ics-eval} first quantifies the query-level disagreement between interleaving-based comparisons and A/B test outcomes described in Section~\ref{sec:iterative_selection}. 
It then evaluates the ability of our Iterative Candidate Selection algorithm to identify the top-$k$ rankers under trust bias while still requiring substantially fewer comparisons than~\citet{bi2023multipleinterleaving}.
For both experiments, we follow the semi-synthetic evaluation protocol of~\citet{brost2016SOSM}, which emulates interleaving interactions using relevance judgments from the MSLR-WEB30K dataset~\cite{qin13MSLR}. 
Each simulation consists of sampling document relevances, generating ranked lists, constructing interleaved lists, and simulating user interactions to infer ranker preferences.

For each query, we uniformly sample a query-document set and randomly select one of the available document features. 
The selected feature is min-max normalized within the query and used as a continuous intrinsic relevance score $rel(i)\in[0,1]$. 
This semi-synthetic evaluation setup is standard in interleaving studies, as no publicly available click datasets provide ground-truth clicks for real-world comparison of interleaving methods.
We then sample ten items with distinct relevance values\footnote{If fewer than ten distinct values are available, a new query-feature pair is sampled. This ensures sufficient variation in relevance for constructing systematically perturbed (worsened) rankings via item-shift described below, although the proposed method does not depend on this constraint.}.

To construct the compared rankers, we first sort the selected items by decreasing relevance, obtaining an ideal ranking. 
$L_2$ is generated by introducing two perturbations, each consisting of shifting an item $i_k$ from position $k$ to a lower position $k_2>k$.
$L_1$ then applies two additional perturbations to $L_2$, making it strictly worse in expectation than $L_2$. 
The interleaved lists are generated according to the approaches described in the following sections.

We simulate user clicks considering two user-behavior models, both subject to trust bias via a term $b(k)=\beta^k$ leading to user clicks following $P(E(i_k)=1|V(i_k)=1) = rel_p(i_k) = \beta^k \cdot rel(i_k)$. 
The first model corresponds to \textit{uniform-relevance users}, who are unable to distinguish differences in document relevance and therefore assign the same intrinsic relevance value to every item, namely $rel(i)=\frac{1}{|I|}\sum_{i^*\in I}rel(i^*)$.
These users effectively click at random, subject only to trust bias. 
Under this setting an unbiased evaluation method should not systematically favor either ranker but, as discussed in previous sections, we expect the method of~\citet{bi2022debiasedinterleaving} to exhibit bias when trust bias is present. 
The second model corresponds to \textit{informed users}, whose engagement probability is proportional to the sampled intrinsic  relevances. 
This setting allows us to evaluate the statistical power of the competing methods.
For both users, examination probability decreases with rank according to the nDCG-inspired decay $V(i_k)=1-\frac{log(k)}{log(k+1)}$ simulating users who may stop browsing before reaching the end of the list $I$~\cite{carterette2011system, bi2022debiasedinterleaving}.

For each experiment, we simulate $n_{users}$ users, each performing $n_{sessions}$ independent search sessions. 
In each session, a new query-document set and a new pair of rankers are sampled.

\subsection{Trust-Aware Balanced Interleaving Bias and Power Evaluation}\label{sec:tabi-eval}
\begin{figure*}[htbp]
    \centering
    \includegraphics[width=0.95\textwidth]{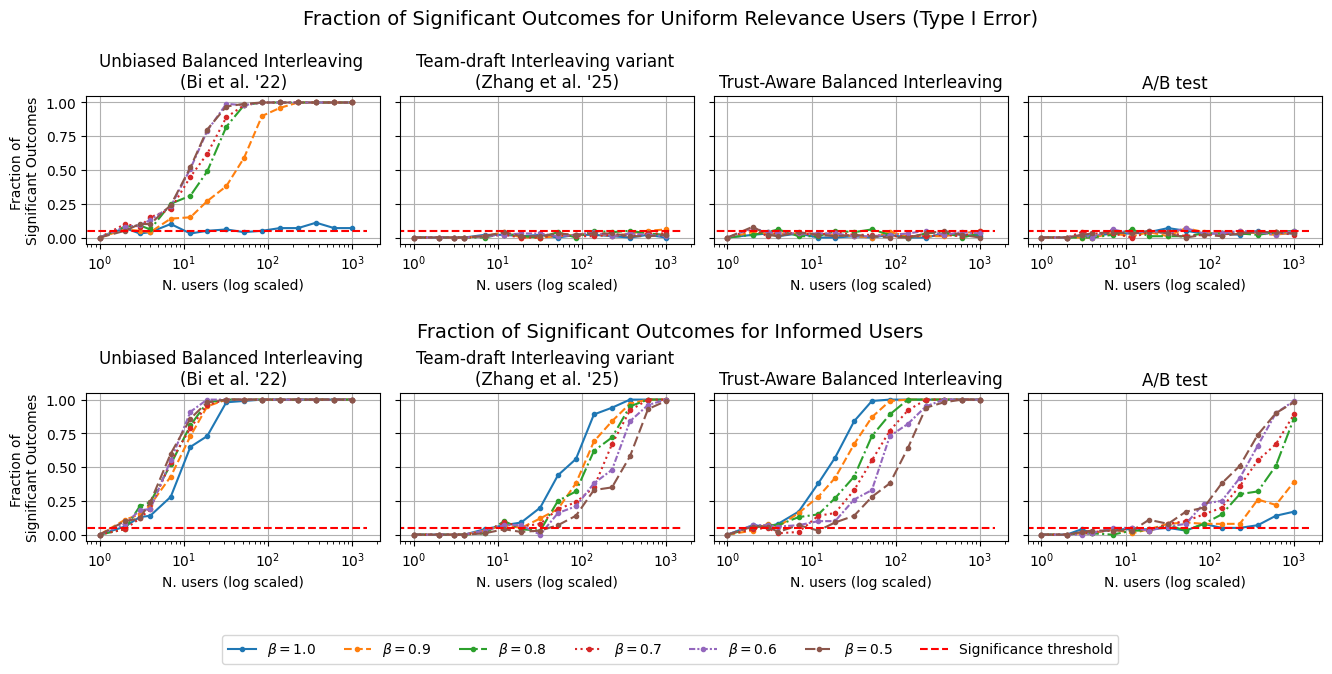}
    \caption{Fraction of significant outcomes produced by the compared online experiments under a uniform-relevance user model (top) and for informed users (bottom), over different trust bias intensities.}
    \label{fig:fig_main}
\end{figure*}
To compare Trust-Aware Balanced Interleaving bias and power against competing interleaving methods, we simulate user interaction under trust bias with strength ranging from $\beta=0.5$ (strong trust bias) to $\beta=1$ (no trust bias).
We run multiple experiments (each repeated 100 times) varying $n_{users}$ and setting $n_{sessions}=10$. 
We compare four evaluation methods: Amazon's approach of~\citet{bi2022debiasedinterleaving}, the team-draft variant of~\citet{zhang2025airbnbTD}, our Trust-Aware Balanced Interleaving (with estimated trust bias $\hat b$), and a standard A/B test comparing the average number of clicks per session generated by each ranking.

Figure~\ref{fig:fig_main} (top) shows the fraction of experiments inferring a statistical significant preference of one ranker over the other from the uniform relevance users.
Since both rankers are equally relevant from the user's perspective, this quantity corresponds to the empirical Type-I error rate. 
As expected, the method of~\citet{bi2022debiasedinterleaving} exhibits systematic bias when trust bias is present, leading to an increasing false-positive rate as the sample size grows. 
This effect disappears when $\beta=1$, confirming that the observed behavior is caused by the interaction between trust bias and positivity violations. 
In contrast, all other methods correctly control the Type-I error rate around the nominal 5\% significance level across all values of $\beta$.

Figure~\ref{fig:fig_main} (bottom) shows the fraction of experiments inferring a statistical significant preference of one ranker over the other from the informed users. 
Trust-Aware Balanced Interleaving and the method of~\citet{bi2022debiasedinterleaving} achieve substantially higher power than the competing approaches, requiring fewer samples to detect differences between rankers. 
We also observe that power decreases as trust bias becomes stronger.
The apparent exception is~\citet{bi2022debiasedinterleaving}, whose power remains artificially high due to the systematic bias highlighted in the previous experiment. 
As shown in Figure~\ref{fig:fig_main} (top), a substantial fraction of these detections correspond to false positives rather than genuine power improvements.

Overall, our Trust-Aware Balanced Interleaving achieves the best trade-off between unbiasedness and fidelity, providing reliable ranker comparisons under trust bias without sacrificing statistical power.
In Appendix~\ref{sec:app_stress_test}, we further stress-test the proposed method by varying the trust-bias function $b$ (including non-decreasing trust biases), the ranked list length $\ell$, and sampling document relevance labels from Yahoo!~\cite{chapelle2011yahoo} and iStella~\cite{dato2016fast} Learning to Rank datasets. 
Across all settings, the results are consistent with those presented above, further supporting our findings.
Moreover, we simulate a scenario in which the trust-bias is estimated in a small-scale experiment prior to the interleaving analysis.
The results show that estimates obtained from as few as $n_{sessions}^{coll}=100$ collection sessions already yield a substantial reduction in bias compared to~\citet{bi2022debiasedinterleaving}.
Moreover, the false-positive rate remains well controlled when $n_{sessions}^{coll}$ is of the same order of magnitude as the number of users participating in the interleaving experiment.

\subsection{Performance of Iterative Candidate Selection}~\label{sec:ics-eval}

\begin{figure*}[th!]
    \centering
    \includegraphics[width=0.95\textwidth]{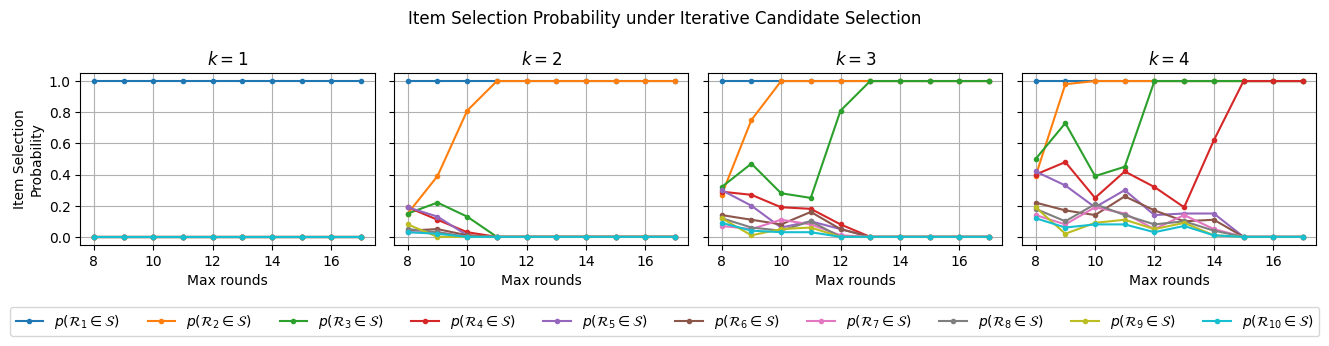}
    \caption{Item selection probability of the Iterative Candidate Selection algorithm for different values of $k$ and $T$.}
    \label{fig:ics}
\end{figure*}

We start by quantifying the extent to which interleaving-based comparisons and standard A/B test outcomes disagree in our setting. 
This analysis is important because the proposed iterative selection strategy relies on interleaving to efficiently compare rankers, and systematic divergence between the two evaluation criteria may affect its ability to reliably retain the best candidates.
These observations motivate the design of our multi-round Iterative Candidate Selection algorithm, whose selection performance under trust bias is evaluated subsequently.

We quantify query-level disagreement using two quantities computed for informed users under no trust bias ($\beta = 1$): (i) expected click-through rate, and (ii) expected credit assignment under different interleaving methods.
A discrepancy is recorded whenever these two criteria induce opposite preferences between $\mathcal{R}_1$ and $\mathcal{R}_2$.
Performing over $100{,}000$ independent runs, we observe a non-negligible disagreement rate of $11.64\%$ for our Trust-Aware Balanced Interleaving and a similar discrepancy ($11.93\%$) for~\citet{bi2022debiasedinterleaving}. 
Estimating expected credit for TDI-based approaches is computationally more demanding, as the interleaving order $o$ may change at each step during the construction of the final ranked list. 
To address this, in each run we approximate the ranker credits by simulating the clicks of one user over 50 user sessions, and then analyze the resulting empirical estimates.
This procedure yields disagreement rates of $30.59\%$ for TDI~\cite{radlinski2008tdi} and $25.87\%$ for its improved variant~\cite{zhang2025airbnbTD}.
These results indicate that disagreement between single sessions of interleaving and A/B metrics is a non-negligible phenomenon not limited to specific classes of interleaving algorithms.
Such behavior persists even under trust-biased user interactions.
Overall, these findings indicate that query-level discrepancies between A/B testing and interleaving-based rankings are substantial, and they may introduce instability in direct-elimination tournament-style selection procedures, motivating our multi-round tournament approach.

We then evaluate our Iterative Candidate Selection algorithm under trust bias ($\beta=0.9$) by simulating a pool of 200 candidate rankers.
For each user session, we assign the list with the highest expected click rate to the ranker with the lowest index, with $\mathcal{R}_1$ denoting the ranker always achieving the highest expected click rate and $\mathcal{R}_{200}$ always the lowest.
We then  run our Iterative Candidate Selection algorithm for $T$ maximum rounds and evaluate its ability to retain the best-performing ranker by estimating the probability that each ranker $\mathcal{R}_i$ is included in the final set of selected candidates $\mathbf{\mathcal{S}}$ over 100 independent runs. 
In each run, we set $\alpha = 0.05$, meaning that each ranker has a $5\%$ probability of a false positive across all $T$ rounds under the null hypothesis, $n_{users}=75$ and $n_{sessions}=10$.
Figure~\ref{fig:ics} reports the selection performance for different values of $k$ and $T$, starting our analysis from $T=8$ rounds, which is the minimum number of elimination rounds required to uniquely identify the best ranker under an infinite-power assumption. 
The experimental results show that our method reliably retains $\mathcal{R}_1$ as the number of rounds increases despite the observed discrepancies between interleaving-based preferences and A/B test outcomes.
Moreover, increasing $T$ improves the recovery of the top-$k$ rankers $\mathcal{R}_1, \dots, \mathcal{R}_k$, while still maintaining its computational efficiency as $T\ll N$.

\section{Conclusions}
In this work, we studied interleaving-based evaluation under realistic user behavior with rank-dependent trust bias. 
While prior work established guarantees under simplified click models, we show that these guarantees can fundamentally break down under realistic trust-biased interactions, revealing a previously unrecognized limitation of existing "unbiased" interleaving methods.
In practice, the presence and strength of trust bias are often unknown and may vary across systems and user populations, making it desirable to adopt evaluation methods that remain valid under both biased and unbiased conditions.

To close this gap, we introduced Trust-Aware Balanced Interleaving, the first interleaving with a credit attribution that provides unbiased comparisons with provable guarantees under trust bias while maintaining fidelity. 
Our analysis further shows that classical unbiased interleaving emerges as a special case of our formulation in the absence of trust bias, allowing us to formally characterize its failure modes.
Importantly, our method does not assume the trust bias to be monotonic, allowing it to capture more general position-dependent trust behaviors. 
This includes vertically ranked interfaces, where users tend to exhibit stronger trust in higher-ranked items~\cite{craswell2008experimental}, as well as more general display layouts such as grid- or card-based interfaces, where trust may depend on spatial presentation rather than a strictly ordered ranking structure~\cite{shen2014webpage, buscher2010good}.

Beyond pairwise evaluation, we investigated the extension of interleaving methods to multileaving settings and identified a fundamental limitation in maintaining fidelity.
This motivated our second contribution, an Iterative Candidate Selection strategy that leverages repeated interleaving comparisons in a Swiss-tournament style. 
The resulting method provides a sub-quadratic alternative to exhaustive pairwise evaluation while maintaining strong empirical agreement with A/B-based ranking outcomes in selecting top-performing systems. 
Importantly, the procedure is simple to implement and inherently interpretable, making it easy to debug and well-suited for deployment in production environments where transparency and reliability of the evaluation pipeline are critical.

From a practical perspective, our results suggest that interleaving-based evaluation can remain both efficient and reliable in operational settings, provided that trust bias is explicitly accounted for and that selection procedures are designed to mitigate potential inconsistencies between interleaving and full A/B testing signals. 
This is particularly relevant in large-scale ranking systems, where the number of candidate models makes exhaustive evaluation infeasible and rapid filtering is essential. 
Importantly, this does not introduce excessive latency overhead in practice, especially when interleaved lists are cached, since Trust-Aware Balanced Interleaving (unlike TDI-based methods) produces only two interleaved lists per pair of rankers, enabling substantial reuse of previously computed interleaved lists for frequent queries.

By providing provable unbiasedness guarantees under realistic user behavior, our framework advances interleaving toward a reliable and principled evaluation paradigm for production ranking systems.
We expect this to significantly compress the runtime footprint of web-scale continuous experimentation engines without sacrificing statistical validity.

\bibliographystyle{ACM-Reference-Format}
\balance
\bibliography{refs}

\newpage
\appendix
\section*{Appendix}
\section{Ex. 1:~\citet{bi2022debiasedinterleaving} biased under trust bias}\label{sec:app_ex_1}
Let us consider a generic trust bias $b(k)$, and $o$ drawn at random following unbiased coin flips. 
Let us also assume for calculation simplicity, that the users see all items of the list, hence $V(i_k)=1$, and note that the example scales trivially if users see only part of the items.
From the credit attribution of Section~\ref{sec:reduction}, in the example of Fig.~\ref{fig:ex_1} we have $\mathbb{E}[\hat d] = 5/7$ for $\mathcal{R} = \mathcal{R}_1$ and $\mathbb{E}[\hat d] = 2/7$ for $\mathcal{R} = \mathcal{R}_2$.
The expected credit attribution of~\citet{bi2022debiasedinterleaving} becomes, in expectation,
\begin{equation}\label{eq:amazon_credit_expectation}
    \mathbb{E}_I[\bar C_\mathcal{R}(I)] = \frac{\mathbb{E}_I[\sum_{i_k\in I} \mathbf{1}(A(i_k=\mathcal{R})E(i_k)]}{\hat d}
\end{equation}
with $E(i_k)=rel(i_k)*b(k)=\alpha*b(k)$.
We then have
\begin{align*}
    \mathbb{E}_I & [\bar C_{\mathcal{R}_1}(I)] = \\ 
    & = \frac{0.5\alpha(b(0)+b(2)+b(3)+b(4)+b(5))}{5/7} +\\ 
    & + \frac{0.5\alpha(b(1)+b(2)+b(3)+b(4)+b(6))}{5/7}
\end{align*}
and
\begin{align*}
    \mathbb{E}_I & [\bar C_{\mathcal{R}_2}(I)] = \\ 
    & = \frac{0.5\alpha(b(1)+b(6))}{2/7}  + \frac{0.5\alpha(b(0)+b(5))}{2/7}
\end{align*}
Therefore the overall credit difference is

\begin{align*}
    \mathbb{E}_I & [\bar C_{\mathcal{R}_1}(I)] - \mathbb{E}_I  [\bar C_{\mathcal{R}_2}(I)] = \\ 
    & = 3.5\alpha[0.4(b(2)+b(3)+b(4))-0.3 (b(0)+b(1)+b(5)+b(6))]
\end{align*}
which is equal to 0 (representing unbiasedness) only in the specific case with $0.4(b(2)+b(3)+b(4))-0.3 (b(0)+b(1)+b(5)+b(6)) = 0$.

\section{Ex. 2: differences in A/B and interleaving results}\label{sec:app_ex_2}
Let us follow the same setup of the previous example, this time having $rel(i_x)=\frac{1}{1+x}$ and taking as bias example $b(k)=0.8^{k-1}$.
The expected clicks for the two versions are
$e_c(\mathcal{R}_1)=\sum_{i_k \in L_1} rel(i_k)*b(k)=1.41$ and 
$e_c(\mathcal{R}_2)=1.36$ resulting the first ranker being better than the second.
Let us now compute the credit attributed by our algorithm to each ranker.
We first calculate $\hat d_{\mathcal{R}_1}^{corr} \approx \frac{2.76}{3.95} \approx 0.7$ and $\hat d_{\mathcal{R}_2}^{corr}\approx \frac{1.19}{3.95} \approx 0.3$.
We then calculate the numerator, that is the expected clicks on each assigned item obtaining (after averaging out the different assignments due to the different draws of $o$) a value of the numerator $n_\mathcal R$ of Eq.~\ref{eq:our_credit_computation} of $\hat n_{\mathcal{R}_1} \approx 0.92$ and $\hat n_{\mathcal{R}_2}\approx 0.44$.
This leads to an expected credit assignment of $\mathbb{E}_I[\bar C_{\mathcal{R}_1}(I)]\approx\frac{0.92}{0.7}\approx 1.31$ and $\mathbb{E}_I[\bar C_{\mathcal{R}_2}(I)]\approx\frac{0.44}{0.3}\approx 1.47$ resulting in the second ranker being better than the first one.

By repeating the same experiment setting $b(k)=0.5^{k-1}$ we observe that the two results agree on $\mathcal{R}_1$ being better than $\mathcal{R}_2$, showing that such difference in results depends both the two ranked lists (with each document relevance) and the trust bias. 
By performing the calculation we have in fact 
$e_c(\mathcal{R}_1)=0.78 > e_c(\mathcal{R}_2)=0.58$, 
$\mathbb{E}_I[\bar C_{\mathcal{R}_1}(I)]\approx\frac{0.51}{0.61}\approx 0.83 > \mathbb{E}_I[\bar C_{\mathcal{R}_2}(I)]\approx\frac{0.15}{0.39}\approx 0.38$.

\section{Pseudocode for Interactive Candidate Selection}\label{sec:app_sel_pseudocode}

\begin{algorithm}
\caption{Interactive Candidate Selection}
\label{alg:interactive_selection}

\KwIn{
Set of rankers $\mathbf{R}$\;
Number of output rankers $k \in \mathbb{Z}^+$\;
Interleaving function $\text{InterleaveFn}(A, B)$\;
Significance threshold $\alpha$\;
Maximum number of rounds $T \in \mathbb{Z}^+$ (default: 8)\;
}

\KwOut{
A list $\mathbf{S}$ containing the top-$k$ rankers\;
}

$\mathbf{S} \gets \mathbf{R}$ \tcp*{Candidate pool}
$stats \gets \{r:\left[0,0,0 \right] \text{ for } r \in \mathbf{R}\}$ \tcp*{Wins, ties, losses per ranker}
$scores \gets \{r : 0 \text{ for } r \in \mathbf{S}\}$\;
$\mathbf{I} \gets \emptyset$ \tcp*{Set of compared ranker pairs}
$t \gets 0$\;

\While{\label{algo:main_loop} $t < T$ \textbf{and} $|\mathbf{S}| > k$}{
    $t \gets t + 1$\;
    $\text{Shuffle}(\mathbf{S})$\;
    $\mathbf{P} \gets \text{DivideInPairs}(\mathbf{S}, \mathbf{I})$\;
    \ForEach{$P \in \mathbf{P}$}{
        \If{$|P| > 1$}{
            $\mathbf{I} \gets \mathbf{I} \cup \{P\}$\;
            $result \gets \text{InterleaveFn}(P[0], P[1], \alpha/T)$\;
            $stats \gets \text{UpdateStats}(stats, result, P[0], P[1])$\;\label{algo:update_counts}
        }
    }    
    $scores \gets \text{CalculateScores}(stats)$\tcp*{Avg. score with wins = 1, ties = 0.5 and losses = 0}
    $currKthScore \gets \text{sorted}(values(scores))[k]$ \tcp*{Decreasing sorting, $k$-highest score}
    \ForEach{$r \in \mathbf{S}$\label{algo:cleanup_for}}{
        $nW, nT, nL \gets stats[r]$\;
        $pVal \gets \sum_{l=nL}^{t} \binom{t}{l} \cdot \left(\frac{1}{3}\right)^{l} \cdot \left(\frac{2}{3}\right)^{t-l}$\;
        \If{$(pVal \le \alpha/kT \textbf{ or } nW + T - t < currKthScore) \textbf{ and } |\mathbf{S}| > k$}{\label{algo:cleanup_line}
            $\mathbf{S} \gets \mathbf{S} \setminus \{r\}$\;\label{algo:cleanup_line_last}
        }
    }
}
$oppStats \gets \{r : 0 \text{ for } r \in \mathbf{S}\}$\;
\ForEach{\label{algo:cleanup_end} $r \in \mathbf{S}$}{
    $\mathbf{O} \gets \text{set of opponents of } r \text{ in } \mathbf{I}$\;
    $oppStats[r] \gets \frac{1}{|\mathbf{O}|}\sum_{o \in \mathbf{O}} score[o]$\;
}
$\mathbf{S} \gets \text{sorted}(\mathbf{S}, score, oppStats)$ \tcp*{Rankers with highest scores first, ties broken by oppStats}\label{algo:sel_sorting}
\Return{$\mathbf{S}[:k]$}\;
\end{algorithm}

Algorithm~\ref{alg:interactive_selection} implements a Swiss-tournament-style candidate selection procedure for identifying a subset of $k$ promising rankers from an initial set $\mathbf{R}$.

At first, the algorithm initializes the candidate set $\mathbf{S}$ with all rankers in $\mathbf{R}$, and creates a statistics dictionary $\textit{stats}$ for each ranker, storing the number of wins, ties, and losses observed during interleaving comparisons. 
An initial score is also assigned to each ranker, and an empty set $\mathbf{I}$ is used to keep track of previously evaluated ranker pairs to avoid redundant comparisons.
The algorithm then proceeds iteratively for at most $T$ rounds, or until only $k$ candidates remain (line~\ref{algo:main_loop}). 
At each iteration, the current candidate set is randomly shuffled and partitioned into disjoint pairs of similar score using $\textsc{DivideInPairs}$, excluding pairs that have already been evaluated.
Each pair of rankers is compared using an interleaving function, which outputs the preferred ranker (or a tie) based on user interaction signals and a statistical test with significance level $\alpha$. 
The outcome of each comparison is used to update the corresponding win--tie--loss counts in $\textit{stats}$ (line~\ref{algo:update_counts}). 
After all pairwise comparisons in a round, the algorithm computes average scores for each ranker based on their accumulated statistics. 
These scores reflect the current evidence of relative performance across all observed comparisons. 
The $k$-th highest score is then used as a dynamic threshold to determine which rankers are still competitive.
Rankers are subsequently filtered (lines~\ref{algo:cleanup_for}-\ref{algo:cleanup_line_last}): those that are statistically significantly outperformed (based on a binomial tail test on losses) or those that cannot reach the current top-$k$ score even under optimistic future outcomes are removed from the candidate set.
This process repeats until either $T$ rounds have been completed or only $k$ rankers remain. 
In the final step (lines~\ref{algo:cleanup_end}-\ref{algo:sel_sorting}), the remaining candidates are ranked according to their final aggregated scores, with ties broken using the average strength of their opponents (opponent scores accumulated through previous comparisons). 
The top-$k$ rankers are then returned in output.

\section{Proof of Theorem~\ref{thm:selection}}\label{sec:app_proof_selection}
\begin{proof}
Let $\mathbf{R}^* \subseteq \mathbf{R}$ be the set of true top-$k$ rankers. 
Given that each interleaving round can have three distinct outcomes (win, tie or loss), we define the null hypothesis $H_0$ such that every true top ranker $r \in \mathbf{R}^*$ has a baseline probability of losing an interleaving match bounded by $P(\text{Loss}) \le 1/3$.
Let $E_{r, t}$ be the event that a true top-$k$ ranker $r \in \mathbf{R}^*$ is erroneously eliminated during round $t \in \{1, \dots, T\}$ via the statistical significance condition of line~\ref{algo:cleanup_line}. 
At any round $t$, the algorithm computes the $p$-value using the cumulative binomial distribution for the observed number of losses $nL$:
\begin{equation*}
    pVal = \sum_{l=nL}^{t} \binom{t}{l} \left(\frac{1}{3}\right)^{l} \left(\frac{2}{3}\right)^{t-l}
\end{equation*}
Because the algorithm triggers a statistical elimination if and only if $pVal \le \alpha/kT$, the probability of committing a Type I error for a single ranker at a single round is bounded by $P(E_{r, t} \mid H_0) \le \alpha/kT$.
The error probability for any item of $\mathbf{R}^*$ to be falsely eliminated at any point during the $T$ sequential rounds can then be recovered by applying the union bound:
\begin{align*}
    P\left(\bigcup_{r \in \mathbf{R}^*} \bigcup_{t=1}^T E_{r, t} \;\middle|\; H_0\right) \le \sum_{r \in \mathbf{R}^*} \sum_{t=1}^T P(E_{r, t} \mid H_0) \\
    \le \sum_{r \in \mathbf{R}^*} \sum_{t=1}^T \frac{\alpha}{k \cdot T}=k \cdot T \cdot \left(\frac{\alpha}{k \cdot T}\right) = \alpha
\end{align*}
\end{proof}

\section{Trust-Aware Balanced Interleaving Extended Results}\label{sec:app_stress_test}
\begin{figure}[htbp]
    \centering
    \includegraphics[width=0.95\linewidth]{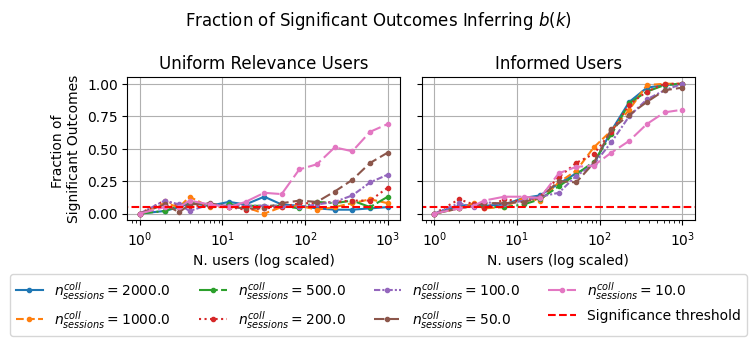}
    \caption{Fraction of significant outcomes produced by each run estimating the trust bias function using $n^{coll}_{\text{sessions}}$ preliminary data collection sessions.}    
    \label{fig:fig_trust_bias}
\end{figure}

\begin{figure*}[h]
    \centering
    \includegraphics[width=0.95\textwidth]{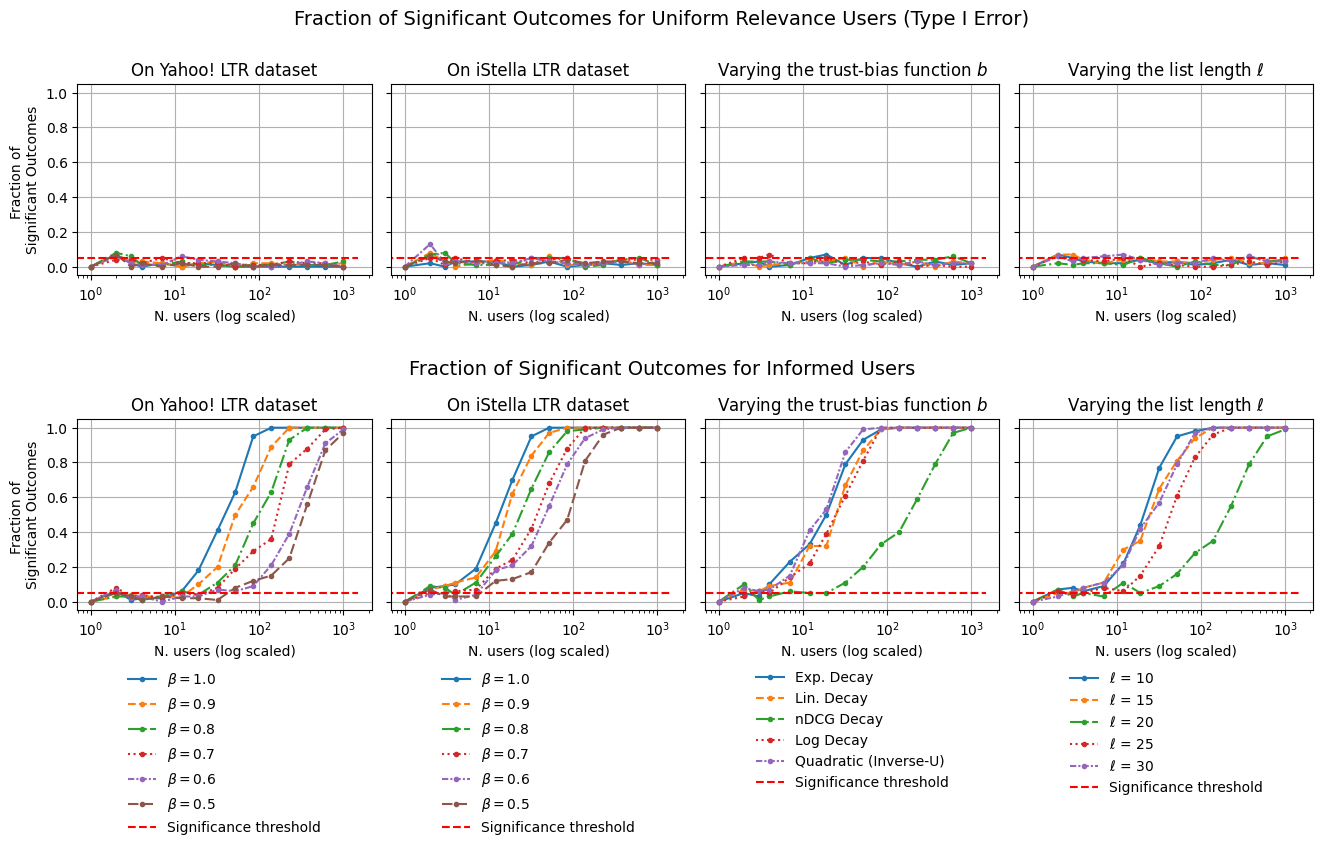}
    \caption{Fraction of significant outcomes produced by each run of Trust-Aware Balanced Interleaving on different setups for uniform relevance users (top) and informed users (bottom).}    
    \label{fig:fig_stress_test}
\end{figure*}

We further stress-tested our unbiasedness claims under a variety of experimental conditions.
Figure~\ref{fig:fig_stress_test} evaluates our Trust-Aware Balanced Interleaving under different setups, both for uniform relevance users (top) and informed users (bottom).
We first repeated the experiment sampling document relevances from the Yahoo! Learning to Rank dataset~\cite{chapelle2011yahoo} (first column) and from the iStella Learning to Rank dataset~\cite{dato2016fast} (second column).
We then tested different trust-bias functions (third column), namely an exponential decay, $b(k)=0.9^{k-1}$, a linear decay,
$b(k)=1-\frac{k-1}{\ell}$, an nDCG-inspired decay, $b(k)=\frac{1}{\log_2(k+1)}$, a logarithmic decay, $b(k)=1/\log(10+k)$, and a reverse-U quadratic function $b(k)=1.5 - \left(2\frac{k}{\ell}-1\right)^2$, which peaks near 1 for central positions and equals 0.5 at the first and last positions.
Lastly, we report the results for different ranked list lengths (last column), $\ell\in\{10,15,20,25,30\}$.
If not otherwise specified, we set $\beta=0.9$.
Across all settings, the results are consistent with those reported in Section~\ref{sec:experiments}, providing further evidence of the unbiasedness of the proposed method.

Lastly, we evaluate the robustness of Trust-Aware Balanced Interleaving when the bias function must be estimated from a data collection experiment of $n^{coll}_{\text{sessions}}$ search sessions. 
During this phase, we record for each rank position $k$ the total number of observed clicks and the expected number of clicks based solely on document relevance, combining the two to estimate the trust bias at position $k$ as $\hat b(k)=N_k^{click}/N_k^{exp}$.
Using the estimated trust bias function $\hat b(k)$, we subsequently run our Trust-Aware Balanced Interleaving on a fresh set of simulated interactions on a different group of user queries computing both the type I error (simulating clicks from uniform-relevance users) and the statistical power (from informed users).
We report results for $\beta = 0.5$; however, the observed trends generalize consistently across other values of $\beta$.
The objective of this experiment is to determine whether Trust-Aware Balanced Interleaving remains robust when the trust bias estimate can be inaccurate due to small sample sizes. 
Figure~\ref{fig:fig_trust_bias} summarizes the results of 100 experiments varying the number of users $n_{users}$ in the second interleaving phase (x-axis) and the number $n^{coll}_{\text{sessions}}$ of data-collection sessions (different lines). 
We observe that accurately estimating the bias function is crucial to preserve the theoretical guarantees of the algorithm. 
As a rule of thumb, we recommend having as many $n^{coll}_{sessions}$ as users in the interleaving experiment.
Despite this requirement, our method remains substantially less biased than~\citet{bi2022debiasedinterleaving}, even with as few as $100$ collection sessions, highlighting its robustness in correcting for trust bias when present.
Finally we observe that, excluding the extreme case of $n^{coll}_{sessions}=10$, $n^{coll}_{sessions}$ does not have a big impact on the fraction of significant outcomes.

\end{document}